# Genetic algorithm enhanced Solovay-Kitaev algorithm for quantum compiling of Fibonacci anyons

Jiangwei Long[1], Xuyang Huang[1], Jianxin Zhong[2,3] and Lijun Meng[1,3,†]

[1] *School of Physics and Optoelectronics, Xiangtan University, Xiangtan 411105, Hunan, People's Republic of China*

[2] *Center for Quantum Science and Technology, Department of Physics, Shanghai University, Shanghai 200444, People's Republic of China*

[3] *Hunan Key Laboratory for Micro-Nano Energy Materials and Devices, Hunan, People's Republic of China*

Quantum compiling, which aims to approximate target qubit gates by finding optimal sequences (braidwords) of basic braid operations, constitutes a fundamental challenge in quantum computing. We develop a genetic algorithm (GA)-enhanced Solovay-Kitaev algorithm (SKA) for approximating single-qubit gates using four elementary braiding matrices (EBMs) derived from Fibonacci anyons. The GA-enhanced SKA demonstrates robust performance, efficiently identifying optimal braidwords within exponentially large search spaces. Notably, the approximation precision achieved by our method surpasses that of Monte Carlo (MC)-enhanced SKA and becomes comparable to deep reinforcement learning (RL) approaches when braidword lengths exceed 25. Implementing 2- and 3-order approximations with the GA-enhanced SKA yields optimal braidword (initial braiding lengths $l_0$=50 and 30 respectively) achieving gate distances of $5.9\times10^{-7}$ - sufficient precision for most quantum computing applications. This work develops an optimized compilation framework for non-Abelian anyon gates, providing an essential methodology for enhancing future topological quantum computation architectures through gate optimization.

## 1. Introduction

The concept of anyons was first theoretically proposed by Myrheim and Leinaas [1] in their seminal 1977 work. Kitaev established the foundational framework for fault-tolerant quantum computation through anyonic systems - now known as topological quantum computing [2]. Topological quantum computing derives its intrinsic fault tolerance from the global topological invariants characterizing anyonic states, which inherently suppress decoherence induced by environmental noise during computational processes[3]. The computational power of topological systems emerges from the braiding statistics of anyons under adiabatic spatial exchanges. While three-dimensional systems exhibit conventional bosonic/fermionic statistics governed by spin, two-dimensional confinement enables the existence of anyonic quasiparticles with fractional statistics. Whereas Abelian anyon braiding induces global phase factors, non-Abelian anyons generate non-trivial unitary transformations in the degenerate ground state manifold - a critical feature enabling universal quantum computation through topological operations. Non-Abelian anyonic systems are exemplified by Ising and Fibonacci anyons. The physical realization of Ising anyons corresponds to Majorana fermions, which are predicted to exist in various physical systems. Specifically, when the *p*-wave superconducting chain model is in a topologically non-trivial phase, localized Majorana zero modes emerge at the two ends of the chain [4]; dipole splitting in p-wave fermionic superfluids gives rise to non-Abelian anyonic excitations [5]x; experimental evidence has demonstrated that superconductor-semiconductor nanowire devices exhibit a robust zero-energy feature [6]; experimental evidence has confirmed the existence of nearly pure Majorana bound states in iron-based superconductors [7]; the thermal Hall conductance of the ν=5/2 fractional quantum Hall state is measured to be $2.5\kappa_0 T\left(\kappa_0=\pi^2 k_B^2/(3h)\right)$, providing evidence for the presence of chiral Majorana fermion edge modes [8]. A critical limitation of Ising anyons lies in their inability to natively implement phase gates through braiding operations alone. Fibonacci anyons represent the simplest non-Abelian system capable of universal quantum computation through braiding operations. While Fibonacci anyons remain experimentally unobserved, theoretical proposals suggest their potential emergence in quantum Hall systems at filling factor ν=12/5 [9], the non-Abelian phase hosting Fibonacci anyons can be realized via uniform interlayer tunneling in the double-layer fractional quantum Hall system [10]. Alternative theoretical frameworks propose Fibonacci anyon behavior in Kondo lattice systems [11]. Topological quantum computation leverages successive braiding operations of non-Abelian anyons, where controlled pairwise exchanges generate unitary transformations through braid matrix multiplication. This process inherently formed the quantum compiling challenge of approximating target quantum gates via braidword optimization. Current compilation strategies encompass hash-based methods [12], algebraic approaches [13,14], RL [15], MC-enhanced SKA [16], and evolutionary algorithms (EA) [17]. While the SKA guarantees universality, its computational efficiency remains suboptimal for large-scale implementations. The integration of EA with the SKA framework (GA-enhanced SKA) significantly improves convergence rates, enabling gate fidelities exceeding 99.9999% through optimized braid sequences of order ≤ 3.

† Corresponding author. E-mail: ljmeng@xtu.edu.cn

Fibonacci anyons are considered the simplest non-Abelian system for realizing universal quantum computation, deriving their name from the fact that the number of possible fusion outcomes grows according to the Fibonacci sequence as the number of anyons $\tau$ increases. This anyonic system exhibits a distinctive fusion rule: $\tau \otimes \tau = \mathrm{I} \oplus \tau$, two Fibonacci anyons $\tau$ fusion yields either another $\tau$ or vacuum I. As illustrated in Fig. 1, single-qubit gate operations are physically realized through the topological manipulation of three Fibonacci anyons, with their computational basis states encoded in the fusion tree Hilbert space.

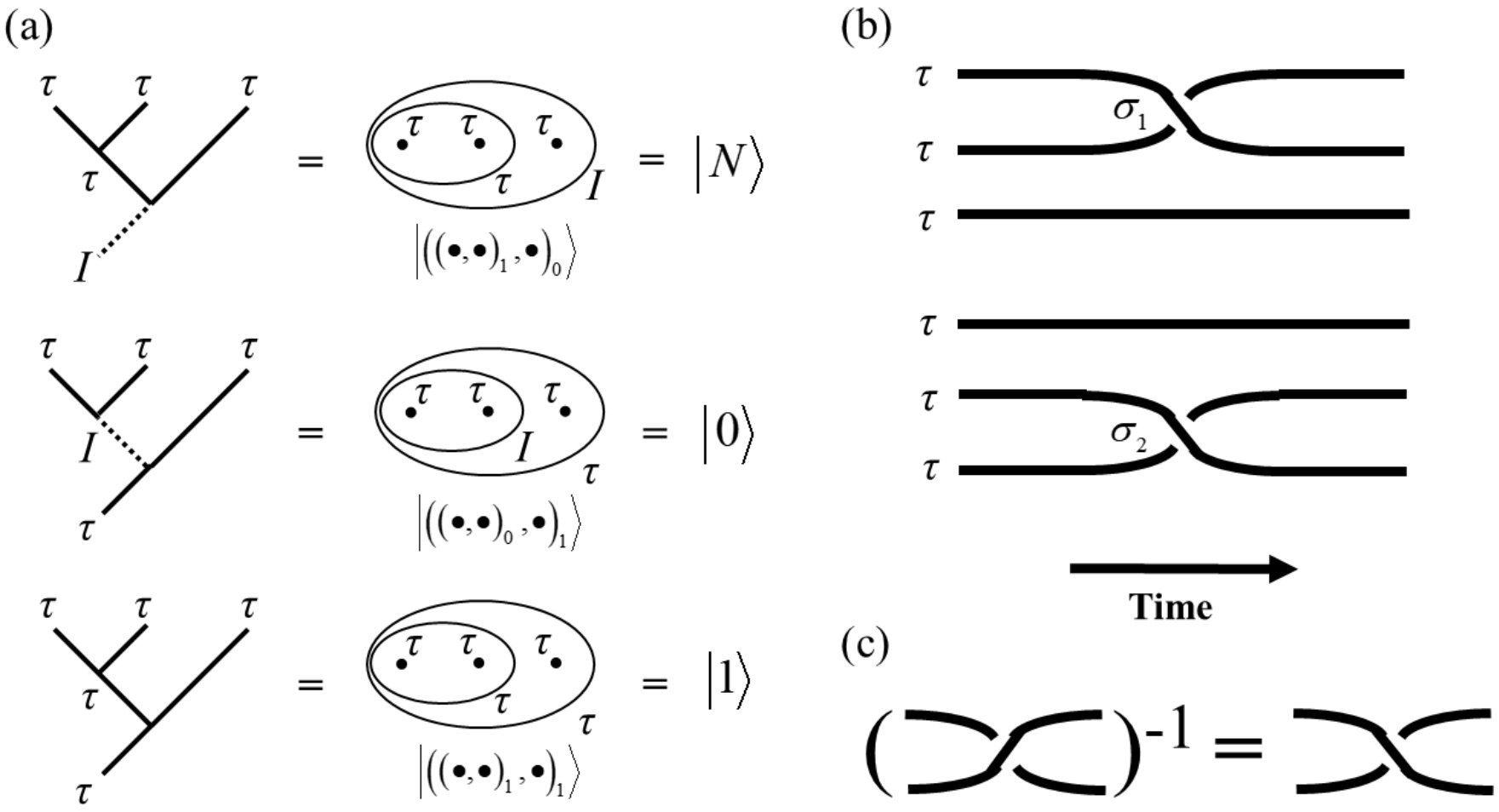

**Fig. 1.** (a) Forming a qubit with three Fibonacci anyons. (b) The two elementary braid operations $\sigma_1$ between anyon 1 and 2 and $\sigma_2$ between anyon 2 and 3. (c) The inverse of the EBM which counterclockwise along the worldline is equivalent to braid clock wisely along the worldline.

Fig. 1 (a) shows that three Fibonacci anyons are used to form a qubit, two $\tau$ fuse into an anyon $\tau$ which then fuses with another $\tau$ into a vacuum state, as a non-computational state $|N\rangle$, is simply labeled as $\left|\left((\bullet,\bullet)_1,\bullet\right)_0\right\rangle$; Two $\tau$ fuse into a vacuum state, and then fuses with another $\tau$ into computational state $|0\rangle$ state, labeled as $\left|\left((\bullet,\bullet)_0,\bullet\right)_1\right\rangle$; Two τ fuse into one τ, and then fuses with another $\tau$ to form $\tau$, forming a computational state $|1\rangle$, labeled as $\left|\left((\bullet,\bullet)_1,\bullet\right)_1\right\rangle$. In Fig. 1(b), $\sigma_1(\sigma_2)$ represents the clockwise exchange of the upper (lower) two anyons along the worldline, and $\sigma_1^{-1}$ or $\sigma_2^{-1}$ represents the counterclockwise exchange of the corresponding two anyons along the world line.

The two basic braid operations $\sigma_1$ and $\sigma_2$ correspond to two EBMs of the form:

$$\sigma_1 = \begin{bmatrix} e^{-i4\pi/5} & 0 \\ 0 & e^{i3\pi/5} \end{bmatrix}, \tag{1}$$

$$\sigma_2 = \begin{bmatrix} -\phi e^{-i\pi/5} & \sqrt{\phi} e^{-i3\pi/5} \\ \sqrt{\phi} e^{-i3\pi/5} & -\phi \end{bmatrix}, \tag{2}$$

$\phi = \dfrac{\left(\sqrt{5}-1\right)}{2}$ in the equation (2), the concrete forms of these two EBMs are derived

from R-move and F-move. The R (rotation) matrix gives the phase factor produced when two Fibonacci anyons move in a certain way towards each other. The F (fusion) matrix is a unitary transformation that maps one of these bases to another. The specific derivation process is suggested in reference [18].

The combination of EBMs $\{\sigma_i^{\pm 1}, i=1,2\}$ enables systematic approximation of arbitrary single-qubit gates through braidword sequences, where gate synthesis is achieved by strategic multiplication of these topological operators in specific orders.

## 1. Genetic algorithm enhanced Solovay-Kitaev algorithm

To quantify the discrepancy between braidword implementations and target single-qubit gates, Hormozi and Zikos [18] along with Bonesteel [19] initially employed the 2 norm for matrix similarity assessment. This approach, however, suffers from sensitivity to global phase factors – quantities that remain physically inconsequential in quantum computational operations. Following established methodologies [14,20], we instead utilize the global phase-invariant error metric $d = d\left(U_0, U\right)$ to evaluate the distance between the unitary realization $U$ (generated by braid sequences) and the target gate $U_0$ (representing ideal single-qubit operations):

$$d\left(U_0, U\right) = \sqrt{1 - \frac{\left|Tr\left(U_0 U^{\dagger}\right)\right|}{2}}, \tag{3}$$

where Tr denotes the trace of $U_0 U^{\dagger}$. When the $U$ approaches the unitary $U_0$, the product $U_0 U^{\dagger}$ approaches the identity such that the error distance $d\left(U_0, U\right)$ tends to 0. The $d\left(U_0, U\right)$ has global phase invariance, that is, the value of the $d\left(U_0, U\right)$ given the $U_0$ is the same for all matrices $U$ up to the global phase change. It is also strictly positive 'exchange' symmetrical $d\left(V, W\right) = d\left(W, V\right)$ and satisfies the triangle inequality $d\left(V, W\right) \le d\left(V, U\right) + d\left(U, W\right)$, where $V$ and $W$ are matrices. When both matrices are multiplied by a coefficient strictly limited to the standard $SU(2)$ form, the second-order norm and the global phase-invariant distance $d$ are actually equivalent, and they differ by a trivial coefficient $\sqrt{2}$ [20]. The $d$ can be effectively reduced by quantum compiling so as to realize a higher precision quantum gate. The central challenge in quantum compiling lies in efficiently identifying braidwords—from an exponentially large configuration space at a fixed length—that exhibit a sufficiently small $d$ relative to a target single-qubit gate.

By employing distance metric $d$ as a similarity measure between matrices and utilizing Genetic Algorithms (GA) to optimize braiding matrix sequences, this approach provides a potential solution framework for quantum compilation challenges involving Fibonacci anyons. The GA, first proposed by John Holland in the 1970s [21], is a method of searching for optimal solutions by simulating natural evolutionary process.

In brief, an initial population of individuals is created, each with a fitness value representing its adaptation to the environment.Through fitness-proportionate selection, individuals demonstrating higher adaptability are retained while less-fit candidates get eliminated. These selected solutions then undergo crossover operations to produce offspring, with a controlled mutation probability introduced to maintain population diversity. The selected subset of offspring with differential evolutionary fitness values constitutes the subsequent population. This iterative selection-crossover-mutation process enables successive generations to significantly enhance population adaptability. GA have been widely employed in physics, facilitating, for example, ab initio prediction of crystal structures and polymorphs without presupposed unit cell parameters or symmetry [22], efficient exploration of alloy atomic configurations guided by target physical properties [23], and implementation of adaptive phase estimation methods for retrieving quantum phases [24]. Theoretical analyses [17] reveal that quantum compilation and GAexhibit structural isomorphism, making GA particularly appropriate for solving quantum compiling challenges. Table 1 systematically demonstrates explicit correspondences between quantum compilation components and GA elements. This alignment enables direct conceptual mapping between the two domains. The distance metric $d$ serves as a functional analog to evolutionary fitness, given their equivalent roles in optimization processes. Solutions exhibiting superior evolutionary fitness persist through iterations, whereas quantum compiling requires identification of braidwords with minimal $d$ relative to target unitary gates. This inverse relationship establishes a critical equivalence: fitness maximization directly correlates with $d$ minimization.

**Table 1** The mapping between quantum compiling and genetic algorithm.

| **Quantum compiling** | | | | |
|---|---|---|---|---|
| Braidword | Thousands of braidwords | Rearrangement braidword | Change EBMs | $d(U_0, U)$ |
| | | ↓ | | |
| **Genetic algorithm** | | | | |
| Individual | Population | Cross-combination | Mutation | Fitness |

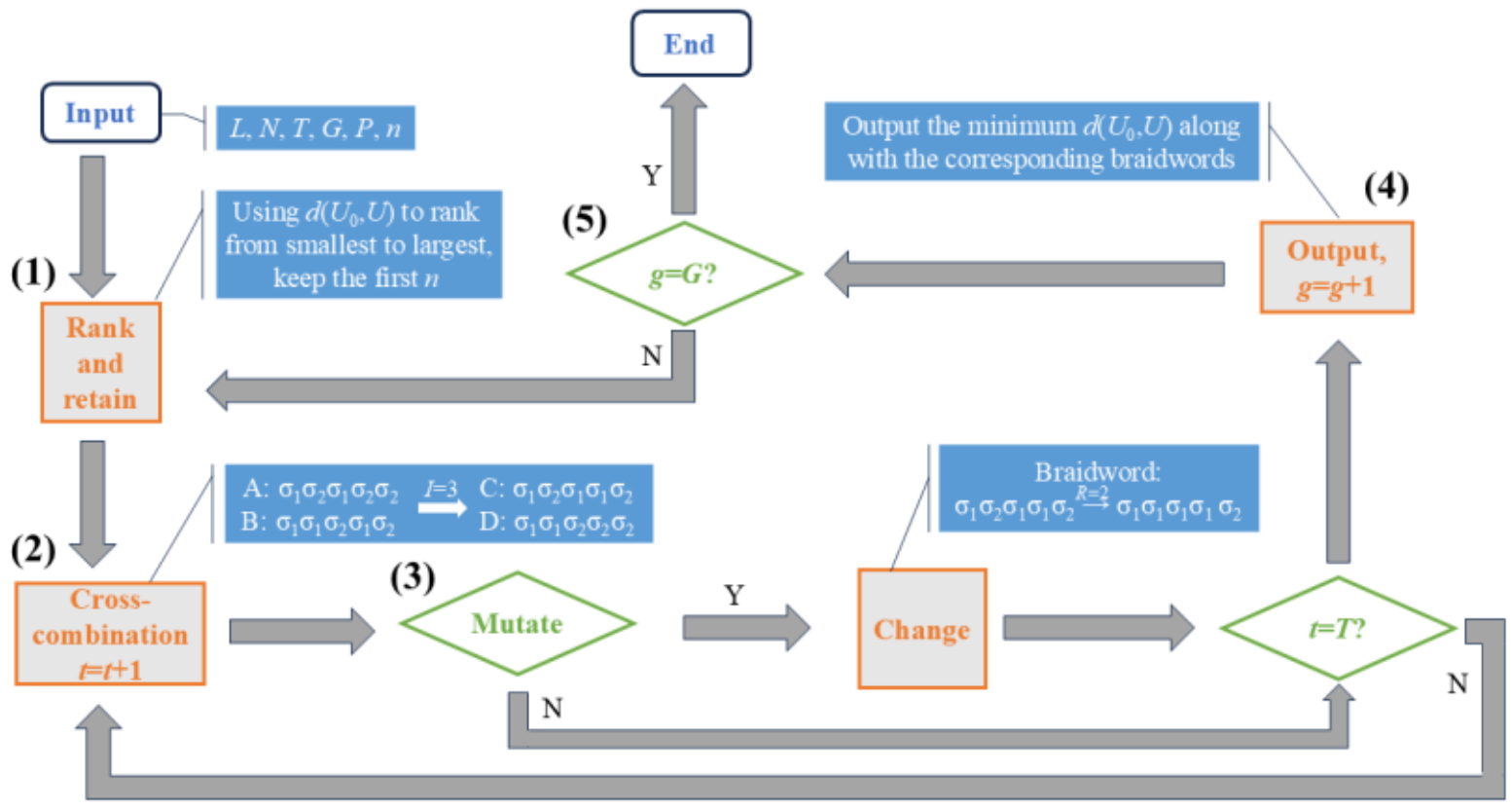

**Fig. 2** The flowchart of GA. The algorithm initializes with input parameters $L$ (the length of the braidword, the number of σ), $N$ (the size of the population, the number of braidwords), $T$ (crossover times), $G$ (maximal generations), $P$ (mutation probability), and $n$ (selection size). Braidword $A$, $B$, $C$, and $D$ are composed of generators $\sigma_i$. In each generation $g$: (1) Braidwords in the population are ranked by $d(U_0,U)$. Top $n$ braidwords are retained. (2) Cross-combination is applied $T$ times to generate offspring. (3) Stochastic alterations are introduced to the braidword through a certain probability $P$. (4) Output the braidword in the offspring with the smallest $d(U_0,U)$ when $t = T$. (5) The algorithm ends when $g = G$.

The GA implementation is formally described in Fig. 2, with core procedures structured as:

(1) Input the standard single-qubit gate that needs to be approximated, the number of EBMs (length, $L$) and types to be used, the number of initial braidwords to be created ($N$), the number of hybridization times ($T = N/2$, performing crossover $N/2$ times yields $N$ new individuals, to maintain a stable population size.), the number of hybridization generation ($G$).The probability ($P$) that a mutation is performed.(2) $N$ initial braidwords of a given length $L$ are randomly generated. The distances $d$ between these braidwords and the target gate are calculated. Subsequently, the top $n$ braidwords (i.e., the selection size) with the smallest $d$ values are selected and retained as the parent pool for crossover in the next generation.

(3) Two braidwords A and B are randomly selected from the parent text, a random integer $I$ (the index of the EBMs in a braidword, counting from left to right) in the range of $0<I\leq L$ is sampled. The subsequence of braidword A from position 1 to $I$ is concatenated with the subsequence of braidword B from position $I$+1 to $L$, forming a new braidword C; the sequence before the $I$ of braidword B and the sequence after the $I$ of braidword A form a new braidword D. Each offspring (C and D) then undergoes mutation with a probability $P$ (the original EBM of a random position $R$ ($0<R\leq L$) in the braidword becomes the EBM of other types).

(4) Repeat (3) so that the number of cross reaches $T$ times. The resulting $N$ offspring, together with the braidword corresponding to the minimum $d$ in the previous generation, are ranked according to the $d$ from smallest to largest, the minimum $d$ in this generation and the corresponding braidword are output. At the same time, the first $n$ are retained as the parent text for the next generation to cross.

(5) Repeat (3) and (4) until the cross generation reaches $G$, stopping the algorithm.

While conventional evolutionary algorithms (EA) employ multi-objective fitness functions incorporating both braid length and $d$ as demonstrated in [17] , our methodology adopts a simplified fitness criterion focusing solely on $d$. This design choice specifically accommodates the fixed-length braid constraint mandated by the SKA.

The SKA promises that for any quantum gate $U$ acting on a single-qubit, it can be efficiently approximated by combining gate operations from a finite set ($\{X, H, T\}$). The pseudocode for SKA is given as follows:

function Solovay-Kitaev$\left(\text{Gate}\, U,\ \text{depth}\ n\right)$

if $\left(n == 0\right)$

  Return Basic Approximation to $U$

else

  Set $U_{n-1} = \text{Solovay-Kitaev}\left(U, n-1\right)$

  Set $V, W = \text{GC-Decompose}\left(UU^{\dagger}_{n-1}\right)$

  Set $V_{n-1} = \text{Solovay-Kitaev}\left(V, n-1\right)$

  Set $W_{n-1} = \text{Solovay-Kitaev}\left(W, n-1\right)$

  Return $U_n = V_{n-1}W_{n-1}V^{\dagger}_{n-1}W^{\dagger}_{n-1}U_{n-1}$

Let $\Delta = UU_0^{\dagger}$, the key of SKA is to do group commutator (GC) decompose, make $\Delta = VWV^{\dagger}W^{\dagger}$. For example, to find a 1-order approximation $U_1$ of $U$, we need to find a 0-order approximation of $V$ and $W$ in the GC decomposition, then $U_1 = V_0W_0V_0^{\dagger}W_0^{\dagger}U_0$, The length of $V_0$, $W_0$ and $U_0$ are all the basic length $l_0$, so the length of each iteration will become 5 times the original length.

$$l_n = l_0 5^n \tag{4}$$

There are three recursive calls per order, so each iteration time becomes:

$$t_n = t_0 3^n \tag{5}$$

For more details on the SKA, we recommend reading the reference [13].

The primary limitation of the SKA arises from the substantial computational cost incurred when using Brute-Force (BF search to derive a 0-order approximation of $U$ for long braidwords, as direct BF search is notoriously inefficient for braidword lengths exceeding 20. The specific procedure of BF search is as follows: For fixed length $L$ of a braidword (number of EBMs), we generate all possible sequences by permuting and combining EBMs selected from the set $\left\{\sigma_1, \sigma_2, \sigma_1^{-1}, \sigma_2^{-1}\right\}$. Then we identify the braidword with the minimal $d$ value among all possible sequences at this $L$. The primary advantage of this method is its guarantee of finding the optimal braidword for the given $L$. However, its major drawback is computational intensity: the number of possibilities requiring exhaustive enumeration scales exponentially as $4^L$, rapidly becoming computationally infeasible for larger values of $L$. In the GA-enhanced SKA, GA replace BF search in each step of the 0-order approximation search, leading to a marked improvement in the SKA's performance. This substitution offers the advantage of enabling an increase in the fundamental length parameter while substantially reducing computational time.

## 2. Results and discussion

The GA requires initialization of critical parameters: population size ($N$=3000) and crossover iterations ($T$=1500).This configuration ensures the number of offspring population is maintained , as each crossover operation between two parent braids produces two descendants ($2\times T$=3000 offspring). The selection derives from parametric analysis in Table 2, demonstrating that maintaining $N$=3000 per generation ($G$) optimally balances algorithmic convergence $d\left(U_0,U\right)$ and computational overhead.

**Table 2.** The convergent $d(U_0, U)$ and time cost for each GA process for the number $N$ of different individuals. The basic length $l_0$ is selected as 50.

| ***N*** | 600 | 1800 | 3000 | 4200 | 6000 |
|---|---|---|---|---|---|
| ***d*(*U*₀,*U*)** | 0.00341693 | 0.00307708 | 0.00192243 | 0.00227884 | 0.00197994 |
| **Time(s)** | 440 | 2062 | 4730 | 8815 | 18124 |

These 3000 offspring and the braidwords with the minimum $d$ in the previous generations are sorted by $d$ in ascending order, we retain the first 2000 offspring for the next generation to cross in order to comply with the principle of the fittest survival in GA. In the offspring generated by each cross, we select the mutation probability $P$ of 0.1 based our consideration of $d$ and computational time as given in Table 3.

**Table 3.** The convergent $d(U_0, U)$ and time cost for different mutation probabilities $P$. The length $L$ was selected as 50.

| ***P*** | 0 | 0.05 | 0.1 | 0.15 | 0.2 |
|---|---|---|---|---|---|
| **$d(U_0,U)$** | 0.00475054 | 0.002816769 | 0.00271369 | 0.00378135 | 0.00324596 |
| Time(s) | 5123 | 5125 | 5064 | 5055 | 5122 |

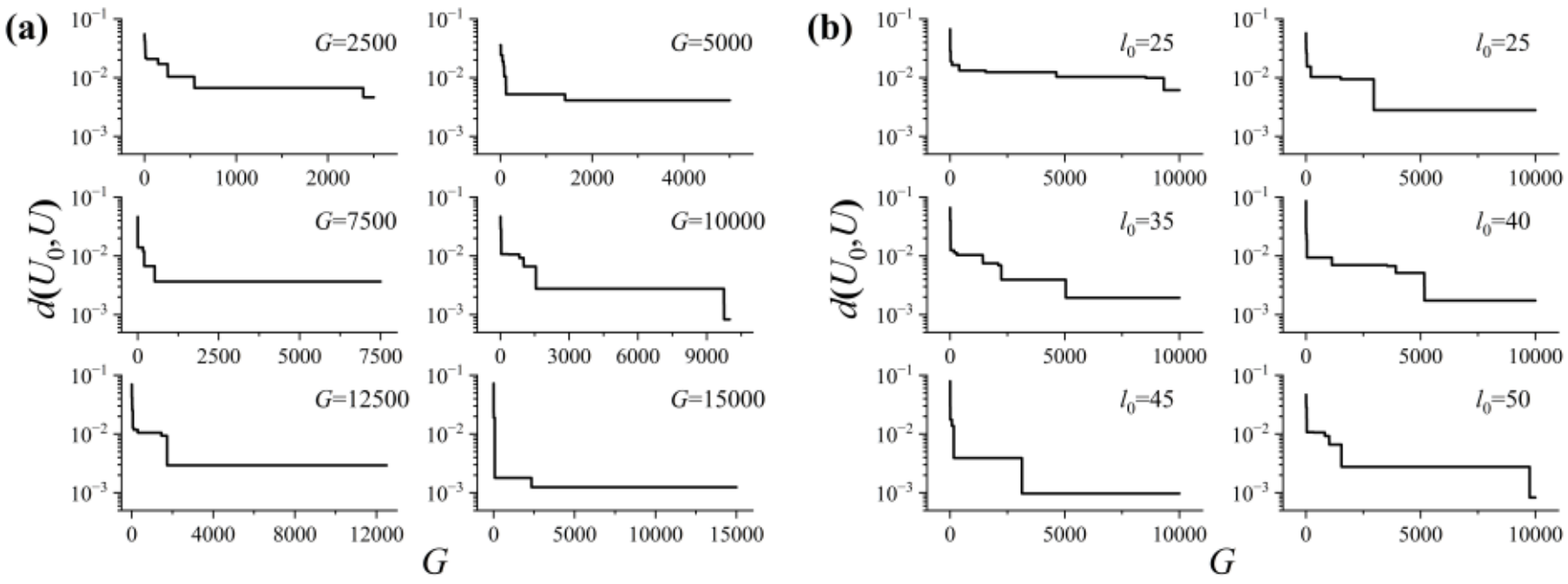

**Fig. 3.** (a) The convergent process of $d$ for different $G$. $l_0$=50. (b) The $d$ as a function of $G$ and $l_0$. The standard single-qubit gate is selected as Hadamard gate.

The number $G$ of hybridization generation is an important parameter which affects the distance $d$ and the Fig. 3(a) give the $d$ as a function of $G$. The Fig. 3(a) shows that, when the $G$ reaches 10000, the distance converges to an ideal value ($<10^{-3}$), and a larger $G$ does not yield higher precision but only increases the time cost. Therefore, in our later calculations, we select the $G$=10000 and obtain optimal results. In Fig. 3(b), GA is used to calculate the 0-order approximation under six basic lengths $l_0$. We find that after 10000 generations ($G$), the distance $d$ decreases in a step-wise manner over approximately 2-5 plateaus before finally converging to a precision between $10^{-3}$ and $10^{-2}$, which is considered to be an ideal 0-order approximation. Compared with the BF search of braidwords of length 20, the $d$ obtained by length 25, 30, 35, 40 is one order of magnitude lower, and 45 and 50 are nearly two orders of magnitude lower. Therefore, the advantage of GA is that the length that can be achieved can easily exceed the BF search of length 20. Time cost in each run is an acceptable time (about 1 hours) for fixed length 20. And more importantly, compared to direct BF search with increase exponentially as the length of braidwords, the time cost in GA increases slightly with the length of braidwords. This will give GA a significant advantage in obtaining longer braidwords.

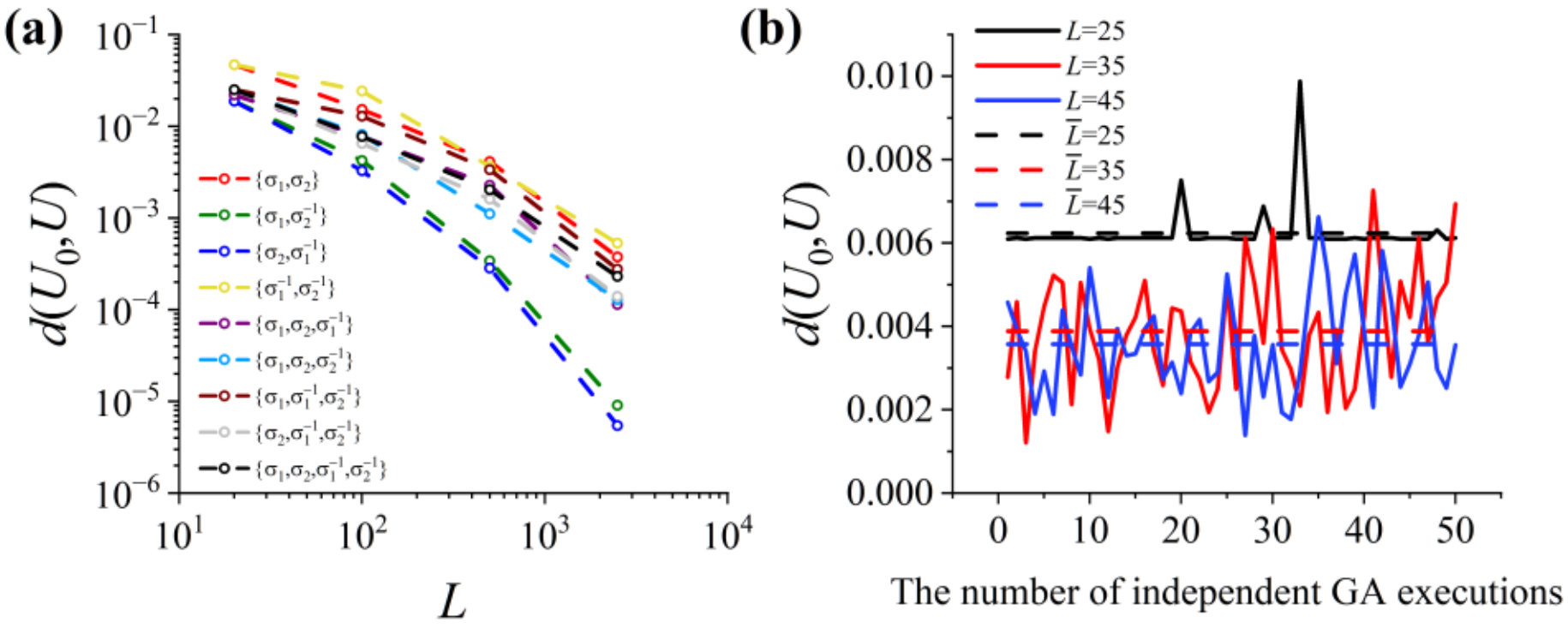

**Fig. 4**: (a) The relationship between the $d$ and the $L$ of braidword at various orders for different types of EBMs adopting GA-enhanced SKA method. The $H$-gate is selected and $l_0$=20. (b) The relationship between $d$ and the number of GA runs for $L$= 25,35,45. The dotted line represents the average $d$ of the corresponding length.

Based on GA-enhanced SKA method, we calculate the $d$ as a function of the length $L$ of braid which is formed by choosing different types of EBM as shown in Fig. 4(a). We find that the $d$ of choosing $\{\sigma_1^{-1},\sigma_2\}$, $\{\sigma_1,\sigma_2^{-1}\}$ at all orders of approximation are lower than that of choosing other types of EBM. Therefore, we choose $\{\sigma_1^{-1},\sigma_2\}$ as EBMs in the subsequent calculations. Note that choosing $\{\sigma_1^{-1},\sigma_2\}$, $\{\sigma_1,\sigma_2^{-1}\}$ is a natural way to avoid the appearance of adjacent pair of $\sigma_1$ and $\sigma_1^{-1}$, or $\sigma_2$ and $\sigma_2^{-1}$ in the braidword, because the multiplication of these two matrices produces the identity matrix, which physically corresponds to a clockwise exchange of two anyons and then counterclockwise exchange along the worldline, and the final result is actually not affected.

As shown in Fig. 4(b), for a short length ($l_0$=25), GA produces relatively stable results as the $d$ when performing multiple calculations, oscillating slightly around 0.006, which indicates that the search performance of GA is strong when the length is near 25, and the resulting minimum value should be approaches to that of BF search. However, at a longer length ($l_0$=35, $l_0$=45), the distance value will fluctuate greatly, most of the $d$ values are between 0.002-0.005, which indicates that GA's performance may have reached its limit after a certain length, which means that when the length is greater than 35, GA's performance will be limited. The dotted line shows that the average $d$ gets smaller as the length gets longer, decreases slightly as the length increases. We set the number of GA runs to multiple times, take the minimum 0-order approximate $d$ found by the GA as the value used in our calculation, this guarantees that the smallest $d$ we can find is most probably around here.

After sifting the above parameters, we get a great set of input parameters, $N$=3000, $T$=$N$/2, $P$=0.1, $G$=10000, the EBM is chosen as $\{\sigma_1^{-1},\sigma_2\}$. Next, we input this set of parameters and use GA-enhanced SKA to find the approximate matrix of single-qubit gate $H$, $X$, $T$.

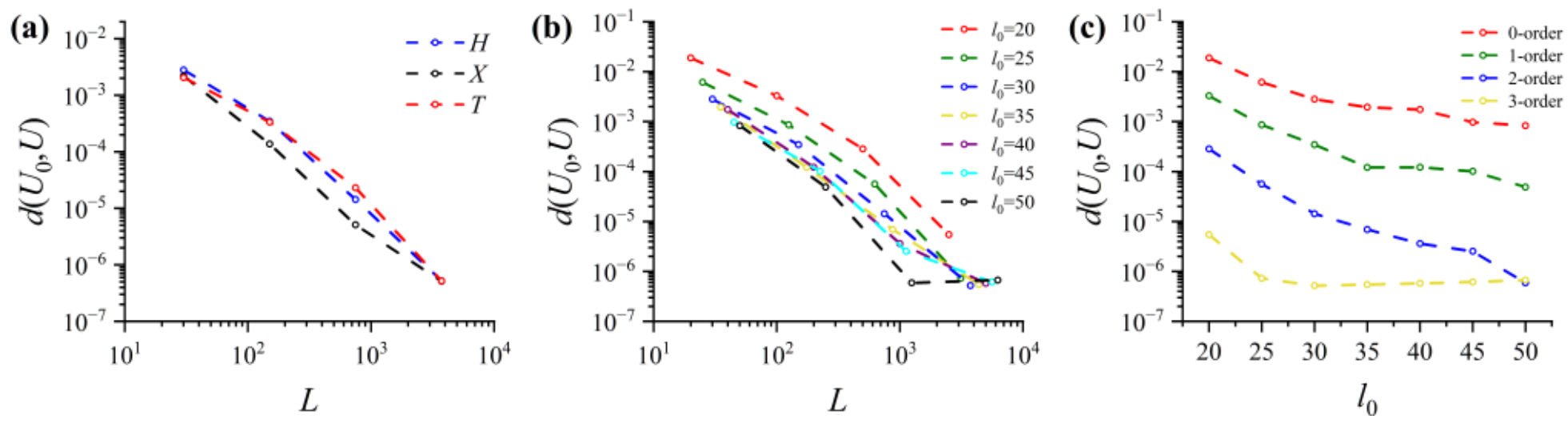

**Fig. 5**: (a) The $d$ of X/H/T gates as a function of length obtained by GA-enhanced SKA. The basic length is 30. (b) The results obtained by selecting different basic lengths and using GA to enhance SKA. (c) The relationship between the 0-, 1-, 2-, 3-order approximation distances of different $l_0$. The standard single-qubit gate is chosen as the $H$.

In Fig. 5(a), the $d$ values for the 0-order approximations show negligible differences across different gates. Although the 1-order and 2-order approximation are different to a certain extent, the 3-order approximation can all reach $d < 10^{-6}$. According to the threshold theorem, an error of less than 1% is acceptable in order to perform a quantum computation with fault tolerance [25,26], which means that the $d$ for the single-qubit gate should be $10^{-3}$. For GA-enhanced SKA, it is easy to find the braidword corresponding to each single-qubit gate. In the Fig. 5(a), the $d$ of the 2-order approximation of the three single-qubit gates are all less than $10^{-3}$, which means that the 2-order approximation is sufficient for the computational requirement in the general case. In Fig. 5(b), we present the various approximation $d$ of braidwords with basic braid lengths $l_0$=20, 25, 30, 35, 40, 45, 50 by GA-enhanced SKA. The trend is that the longer the basic length is chosen, the smaller the approximate distance $d$ of each order will tend to be. As shown in Fig. 5(c), the approximation error $d$ for each order decreases as $l_0$ increases. The error values for different approximation orders differ by approximately one order of magnitude, except for the 3-order case.We notice that in the 3-order approximation in the fig.5(c), when the $l_0$ is 20, 25, and 30, the $d$ decreases continuously, but when the $l_0$ is 35,40,45,50, the $d$ increases slightly, which means that there is a critical length at which there is a minimum $d$ for the 3-order approximation. The minimum $d$ should be slightly less than the 3-order approximate distance ~$5.2\times10^{-7}$($5.1994163\times10^{-7}$) with basic length of 30. The resulting braidwords achieve a significantly lower $d$ to a standard single-qubit gate than both conventional SKA and MC-enhanced SKA methods. This advancement holds considerable implications for high-fidelity fault-tolerant quantum computation. Even if the $l_0$ is further increased, the $d$ should not decrease significantly. We note that when the $l_0$ is 50, the 3-order approximation does not become smaller than the 2-order approximation, which implies that the $d$ has reached its minimum in the 2-order approximation, and it is difficult to continue to reduce $d$ (i.e. to improve precision). In our calculations, the 2-order approximate $d$ with the basic length of 50 is ~$5.9\times10^{-7}$($5.8854870\times10^{-7}$). Indeed, it is close to a 3-order approximate $d$ of basic length 30. At the same time, the $d$ of the 2-order approximation with an $l_0$ =50 is close to the $d$ of the 3-order approximation of other basic length. According to the Eq. (2), the length $L$ corresponding to the 2(3)-order approximation with a basic length $l_0$ of 50(30) is $l_2$=1250($l_3$=3750). Then we choose 2-order approximation with $l_0$ =50, the required braiding operations will be greatly reduced (here from 3750 to 1250), which means that we can simplify operations and save time by choosing a lower order approximation with a longer basic braid length.

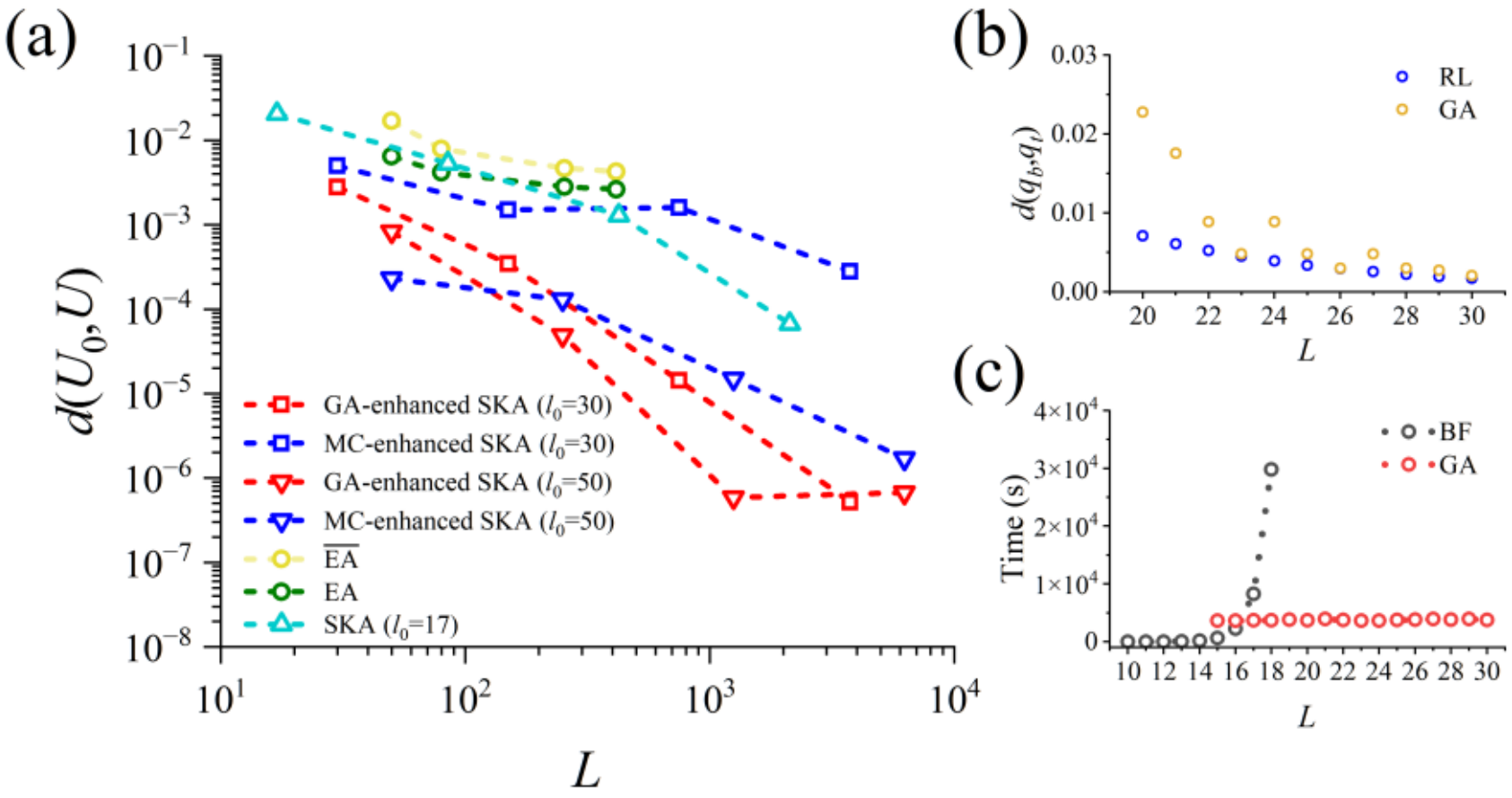

**Fig. 6**: (a) The performance comparison of EA, SKA, GA-, and MC-enhanced SKA. (b) The performance comparison between RL and GA. (c) The comparison of BF and GA run time at different lengths of braidword.

In our GA, the $G(N)$ are 10000(3000), far exceeding the 400(80) of the EA, this results in an improved performance. In Fig. 6 (a), we compare the performance of GA with EA, the distances $d$ of 0-order approximation GA are smaller than those of EA for both average (yellow line) and minimum (green line) values by running 100 times [17]. The standard SKA , GA- , MC-enhanced SKA are also compared in Fig. 6 (a), here the MC-enhanced SKA data comes from the reference [16]. Fig. 6 (c) compares the runtimes of GA and BF across varying lengths of braidword. Note that the evaluated length of braidword ranges is 15-30 for GA and 10-18 for BF. As evident from Fig. 6 (c), GA runtime remains stable (fluctuating between ~3600-4000 s) despite increasing the length of braidword. This aligns with our prior analysis: GA execution time depends primarily on crossover operations (population size) and the number of offspring generations. In contrast, BF exhibits exponential runtime growth due to the combinatorial explosion of exhaustive possibilities. At length 17, BF requires 8260.5 s—exceeding GA's 3720.4 s. This divergence amplifies at length 18 (BF: 29827.0 s and GA: 3700.5 s). For standard SKA implementations, we selected a base length $l_0$=17, because we try to demonstrate GA-enhanced SKA's superiority over standard SKA, BF's computational cost (8260.5 s) remains acceptable while exceeding GA's runtime. As shown in Fig. 6(a), the GA-enhanced SKA (red line) achieves significantly lower $d$ values at each approximation order compared to standard SKA (light blue line). Notably, the execution time to get 3-order approximation for standard SKA is approximately $8260\times3^3$ s, while GA-enhanced SKA requires only ~$3800\times3^3$ s. This represents a substantial reduction in total computation time and lower $d$ significantly, demonstrating that GA-enhanced SKA outperforms standard SKA. Note that, in the MC algorithm, four EBMs map to four spin states $\left(|\uparrow\rangle,|\downarrow\rangle,|\leftarrow\rangle,|\rightarrow\rangle\right)$, a braidword to a one-dimensional spin chain. Starting from a braidword composed of four EBMs with equal possibility, the type of each EBMs changes randomly from the left end to the right end in a braidword. If the change causes the $d$ to become smaller, it will directly be accepted and move to the next position, otherwise accept this state with probability $p=e^{-\left(E_{\uparrow}-E_{\downarrow}\right)/T_\alpha}$ , $E\uparrow$ ($E\downarrow$) and $T_\alpha$ represent the energy of spin up(down) and dimensionless temperature. Then the position of change move along spin chain, so as to obtain smaller and smaller $d$ [16]. When the basic length $l_0$ is 30, we observe the performance of each order approximation of GA-enhanced SKA is better than that of MC-enhanced SKA. When the $l_0$ increases to 50, the 1-, 2- and 3-order approximations obtained by GA-enhanced SKA are superior to those obtained by MC-enhanced SKA. The recently developed RL method is also an excellent approach to solve topological quantum compiling problem [15]. This approach relies on policy network and target network, and using $f(s)=\lambda G(s)+J(s)+d(s)$ as evaluation function, in this function, $G(s)$ is the actual cost from the initial state $s_i$ (an old braidword) to the current state $s$ (a new braidword) , and $\lambda\in[0,1]$ is a weighting factor, $J(s)$ is cost-to-go function, $d(s)$ is a decimal-penalty term. To measure the accuracy of the braidword, Y H Zhang et al. use the quaternion distance $d(q_b,q_t)=\sqrt{1-\langle q_b,q_t\rangle}$, where $q_b$ and $q_t$ are the unit quaternions corresponding to the unitary from the braidword and the target gate, respectively, and $\langle q_b, q_t \rangle$ denotes their inner product [27]. We compare the performance of GA and RL by $d(q_b, q_t)$ for braidwords with length 20~30, and present the results in Fig. 6 (b). The data of RL comes from the fitting formula $L = 1.55\log(1/\varepsilon)^{1.6}$ $(\varepsilon = d(q_b, q_t))$ in the literature [15]. Although the performance of GA is worse than that of RL at lengths of 20 to 25, the $d(q_b, q_t)$ obtained by the two methods can be compared with each other at lengths greater than 25. At shorter lengths, the $d(q_b, q_t)$ obtained by using GA-enhanced SKA is larger, the possible reason is that only two kinds of EBMs were selected. Another advantage of this GA is that it takes less time. It will take several days to perform a BFsearch with a length of 13 (a BF search with a length exceeding 13 becomes infeasible), but for GA, we have a huge advantage in terms of time. The running time of the algorithm is about 1 hours, and it only takes about one day to complete the 3-order approximation with GA-enhanced SKA, and the increase in length will not bring significant increase in time cost.

## 3. Conclusion and perspectives

This work establishes a GA-enhanced SKA approach for Fibonacci anyon quantum compilation, demonstrating strong performance. The performance of the GA-enhanced SKA, which involves many adjustable parameters (e.g., *P*, *N*), can be improved (leading to better results or faster convergence) by optimizing these parameters. The mutation probability *P* of 0.05~0.1 and the number of initial braidwords *N* of 1800~4200 are chosen for the compiling of single-qubit quantum gates. We calculate the various order approximations of three common single-qubit gates (*X*, *H*, *T*) and evaluate the performance of GA-enhanced SKA by comparing it with MC-enhanced SKA and deep reinforcement learning. For single-qubit gates, GA-enhanced SKA with basic braiding length $l_0$=30, the 3-order approximation gives an approximate braidword with distance ~$5.9\times10^{-7}$. Furthermore, for $l_0$=50, a 2-order approximation can produce a braidword with comparable precision, which can greatly reduce unnecessary braiding operations. The achieved precision of single-qubit quantum gate is superior to that of the MC-enhanced SKA and comparable to that of the deep RL method, being sufficient for most quantum computing applications. Our work enriches the set of approaches to solving and optimizing quantum compilation problems and will be useful for topological quantum computation in the future.

## Acknowledgments

We wish to thank Yizhi Li for assistance in revising this article and the insightful discussions about comparison of time performance between GA and BF search.

This work is supported by the National Natural Science Foundation of China (Grant Nos. 12374046, 11204261), College of Physics and Optoelectronic Engineering training program, a Key Project of the Education Department of Hunan Province (Grant No. 19A471), Natural Science Foundation of Hunan Province (Grant No. 2018JJ2381), Shanghai Science and Technology Innovation Action Plan (Grant No. 24LZ1400800), Education Department of Hunan Province (Grant No. 24C0316).

**Author Contributions** Jiangwei Long and Xuyang Huang are co-first authors of this article. Jiangwei Long: Conceptualization, Investigation, Methodology, Writing (original draft), Validation. Xuyang Huang: Methodology, Software. Jianxin Zhong: Supervision, Funding acquisition, Project administration, Resources, Writing (review and editing). Lijun Meng: Supervision, Funding acquisition, Project administration, Resources, Writing (review and editing).

**Data availability statement**

The data that support the findings of this study are available upon reasonable request from the authors. The implementation code for genetic algorithms and group commutator decomposition presented in this work has been made publicly available on the GitHub repository: https://github.com/LongJiangwei/code python_code.zip